\documentclass[aps,showpacs,prd,twocolumn]{revtex4}
\usepackage{graphicx}
\usepackage{amssymb}
\usepackage{amsmath}

\newcommand{\be}{\begin{equation}}
\newcommand{\ee}{\end{equation}}
\newcommand{\ben}{\begin{eqnarray}}
\newcommand{\een}{\end{eqnarray}}
\newcommand{\bes}{\begin{subequations}}
\newcommand{\ees}{\end{subequations}}

\begin{document}

\title{Twinlike Models for Self-Dual Maxwell-Higgs Theories}
\author{D. Bazeia$^{1,2}$, E. da Hora$^{1,3}$ and R. Menezes$^{2,4}$.}
\affiliation{$^{1}${Departamento de F\'{\i}sica, Universidade Federal da Para\'{\i}ba,
58051-900, Jo\~{a}o Pessoa, Para\'{\i}ba, Brazil.}\\
$^{2}${Departamento de F\'{\i}sica, Universidade Federal de Campina Grande,
58109-970, Campina Grande, Para\'{\i}ba, Brazil.}\\
$^{3}${Department of Mathematical Sciences, Durham University, DH1 3LE,
Durham, County Durham, U.K.}\\
$^{4}${Departamento de Ci\^{e}ncias Exatas, Universidade Federal da Para%
\'{\i}ba, 58297-000, Rio Tinto, Para\'{\i}ba, Brazil.}}

\begin{abstract}
In this work we present a theoretical framework that allows for the
existence of coherent twinlike models in the context of self-dual
Maxwell-Higgs theories. We verify the consistence of this framework by using
it to develop some twinlike self-dual Maxwell-Higgs models. We use a
combination of theoretical and numerical techniques to show that these
models exhibit the very same topological BPS structures, including their
field configurations and total energy. The study shows that it is possible to
develop a completely consistent prescription, which extends the idea of
twinlike models to the case of vortices in Maxwell-Higgs theories.
\end{abstract}

\pacs{11.10.Kk, 11.10.Lm}
\maketitle


\section{Introduction}

Topological structures have been extensively used to study phenomena
related to several areas of physics \cite{n5}. In
particular, they are of direct interest to cosmology, since they can be formed
in a rather natural way during phase transitions in the early universe.
In the context of classical field theories, topological structures are
described as finite energy solutions to some nonlinear field
models. Usually, the models are endowed with a scalar potential
which exhibits degenerate vacua, allowing for the spontaneous
symmetry breaking mechanism. In this context, kinks \cite{n0}, which are the
simplest topologically non-trivial configurations, are the 1-dimensional
solutions to a single real scalar field theory, while vortices \cite{n1} and
monopoles \cite{n3} are, respectively, 2-dimensional solutions to some
Abelian gauge theories and 3-dimensional solutions to non-Abelian gauge
theories.

An interesting issue concerning the study of topological structures is that,
in some particular cases, such structures, beyond solving the standard
equations of motion, are also solutions to a set of coupled first-order
differential equations, named Bogomol'nyi-Prasad-Sommerfield (BPS) equations 
\cite{Bogo}. In this context, since we do not need to deal with the
second-order Euler-Lagrange equations of the model under investigation, the
topologically non-trivial BPS configurations are easier to be obtained.
Another advantage of studying BPS configurations is that they describe minimum
energy solutions to the corresponding classical field theories.

Motivated by recent results on cosmology and superstring
theories, a new type of classical field theories has been intensively
investigated during the last years. These theories, named \textit{k-field}
models, are usually endowed with non-standard kinetic terms that 
change the dynamics of the model under investigation. In this
context, topologically non-trivial BPS solutions may exist, and the
features they engender can be quite different, or quite similar to the ones
presented by their usual counterparts; see, for instance, Ref.~\cite{todos}.

In a recent work, Andrews, Lewandowski, Trodden and Wesley noted that some
k-theories can mimic the very same topological structures engendered by
their standard counterparts, including their energy densities \cite{ALTW}.
Since these models map each other solutions, they were named \textit{twinlike models}.
In another recent work, some of us studied the existence of twinlike models
in the context of scalar field theories \cite{ufpb}. There, in
particular, one used a combination of theoretical and numerical techniques
in order to perform a detailed analysis of such models, including the
presence of first-order differential equations and their applications to the
braneworld scenario. Soon after, another work by Adam and Queiruga \cite%
{adam} made very good use of the first-order framework, to develop algebraic
technique to construct twinlike theories. Yet more recently, another
interesting result was introduced in Ref.~\cite{bm}, concerning the presence of twinlike
models, where the models present not only the same solution with the same
energy density, but also the very same stability features.

The above works on twinlike models deal with scalar fields. Thus, a rather
natural question is how to construct consistent twinlike models under the
action of gauge fields. In the present work we focus our attention on planar
Maxwell-Higgs models and their topologically non-trivial self-dual
solutions. To achieve this goal, we organize the work as follows: in the
next section we introduce the generalized theory and we
also present the theoretical framework which allows for the existence of
twinlike self-dual Maxwell-Higgs models. In Sec.~ \ref{numerical}, we prove
the consistency of such framework by using it to developed a few twinlike
self-dual Maxwell-Higgs systems. Next, in Sec.~\ref%
{numerical2} we perform a detailed numerical analysis of these systems and
 in Sec.~\ref{end}, we present conclusions and some perspectives
concerning future investigations.


\section{The theoretical framework}

\label{general}

Let us first present the model, which is defined by the planar Lagrange
density%
\begin{equation}
\mathcal{L}=-\frac{h\left( \left\vert \phi \right\vert \right) }{4}F_{\mu
\nu }F^{\mu \nu }+w\left( \left\vert \phi \right\vert \right) \left\vert
D_{\mu }\phi \right\vert ^{2}-V\left( \left\vert \phi \right\vert \right) 
\text{ .}  \label{1}
\end{equation}%
Here, $F_{\mu \nu }=\partial _{\mu }A_{\nu }-\partial _{\nu }A_{\mu }$
stands for the electromagnetic field strength tensor, $D_{\mu }\phi
=\partial _{\mu }\phi +ieA_{\mu }\phi $ represents the usual covariant
derivative and $V\left( \left\vert \phi \right\vert \right) $ is the
spontaneous symmetry breaking potential which controls the scalar matter
self-interaction. Also, $h\left( \left\vert \phi \right\vert \right) $ and $%
w\left( \left\vert \phi \right\vert \right) $ are dimensionless functions of
the amplitude of the scalar field. We note that non-trivial choices to $h\left( \left\vert \phi
\right\vert \right) $ and $w\left( \left\vert \phi \right\vert \right) $
will induce, respectively, non-standard dynamics to both the gauge and
scalar fields. Also, there are interesting
motivations concerning the use of these two functions and, in particular, the function 
$h\left( \left\vert \phi \right\vert \right) $ represents a \textit{%
generalized dielectric function}, which can be used to describe
interactions between quarks and gluons \cite{quarks}.

Since it is useful to deal with dimensionless fields, coordinates and parameters, let us introduce 
the mass scale $M$, which we use to implement the following
scale transformations: $x^{\mu }\rightarrow x^{\mu }/M$, $\phi \rightarrow
M^{\frac{1}{2}}\phi $, $A^{\mu }\rightarrow M^{\frac{1}{2}}A^{\mu }$, $%
e\rightarrow M^{\frac{1}{2}}e$ and $\upsilon \rightarrow M^{\frac{1}{2}%
}\upsilon $, where $\upsilon $ represents the spontaneous symmetry breaking
parameter of the model under investigation. As a consequence, we get that
$\mathcal{L\rightarrow }M^{3}\mathcal{L}_{g}$, where $\mathcal{L}_{g}$
stands for the dimensionless Lagrange density to be
used from now on, which has the very same functional form of $\mathcal{L}$.
Also, for simplicity, we choose $\upsilon =e=1$.

The standard procedure to search for the rotationally symmetric
configurations is to consider such configurations as static solutions of the
Euler-Lagrange equations of the model. In the present case, these equations
are%
\begin{equation}
\partial _{\mu }\left( hF^{\mu \nu }\right) =J^{\nu }\text{ ,}  \label{2}
\end{equation}%
\begin{eqnarray}
&&\left. \partial _{\mu }\left( w\partial ^{\mu }\left\vert \phi \right\vert
\right) -A_{\mu }A^{\mu }\left\vert \phi \right\vert w=\frac{1}{2}\left\vert
D_{\mu }\phi \right\vert ^{2}\frac{dw}{d\left\vert \phi \right\vert }\right.
\;\;\;\;\;\;\;\;\;\;\;  \notag \\
&&\left. \text{ \ \ \ \ \ \ \ \ \ \ \ \ \ \ \ \ \ \ \ \ \ \ }-\frac{F^{2}}{8}%
\frac{dh}{d\left\vert \phi \right\vert }-\frac{1}{2}\frac{dV}{d\left\vert
\phi \right\vert }\text{ ,}\right.  \label{3}
\end{eqnarray}%
where $F^{2}=F_{\mu \nu }F^{\mu \nu }$. Also, we take $J^{\mu
}=-2w\left\vert \phi \right\vert ^{2}A^{\mu }$.

In the present case, the Gauss law for static fields can be written in the
form ($k$ represents spatial indices)%
\begin{equation}
\partial _{k}\left( h\partial ^{k}A^{0}\right) =-2w\left\vert \phi
\right\vert ^{2}A^{0}\text{ ,}  \label{4}
\end{equation}%
from which we note that such law is trivially satisfied by $A^{0}=0$
(temporal gauge). So, we fix this gauge and use it from now on.

We now look for static and rotationally symmetric configurations of the form%
\begin{equation}
\phi \left( r,\theta \right) =g\left( r\right) e^{in\theta }\text{ ,}
\label{5}
\end{equation}%
\begin{equation}
\mathbf{A}\left( r,\theta \right) =-\frac{\widehat{\theta }}{r}\left(
a\left( r\right) -n\right) \text{ ,}  \label{6}
\end{equation}%
where $r$ and $\theta $ are the polar coordinates. Also, $n=\pm 1,$ $\pm 2,$ 
$\pm 3,...,$ represents the vorticity of the configuration. We use the above
 eqs.~\eqref{5} and \eqref{6} into the Euler-Lagrange equations %
\eqref{2} and \eqref{3} to get
\begin{equation}
h\left( \frac{d^{2}a}{dr^{2}}-\frac{1}{r}\frac{da}{dr}\right) +\frac{da}{dr}%
\frac{dh}{dr}=2wg^{2}a\text{ ,}  \label{7}
\end{equation}%
\begin{eqnarray}
&&\left. w\left( \frac{d^{2}g}{dr^{2}}+\frac{1}{r}\frac{dg}{dr}-\frac{a^{2}g%
}{r^{2}}\right) =\frac{1}{4}\left( \frac{1}{r}\frac{da}{dr}\right) ^{2}\frac{%
dh}{dg}\right.  \notag \\
&&\left. \text{ \ \ \ }-\frac{1}{2}\left( \left( \frac{dg}{dr}\right) ^{2}-%
\frac{g^{2}a^{2}}{r^{2}}\right) \frac{dw}{dg}+\frac{1}{2}\frac{dV}{dg}\text{
.}\right.  \label{8}
\end{eqnarray}%
These are the rotationally symmetric equations of motion for the fields $%
a\left( r\right) $ and $g\left( r\right) $, respectively.

In order to solve the Euler-Lagrange equations \eqref{7} and \eqref{8},
we need to specify the model. In general, it can be done
following the generalized model, i.e., choosing non-trivial functional forms
to the functions $h\left( g\right) $ and $w\left( g\right) $. In this case,
it is important to note that both functions must be positive-definite,
in order to avoid problems with the positiveness of the total energy of the
model; see the expression for the energy density below. We have to choose $%
h\left( g\right) $ and $w\left( g\right) $, and we also need to specify the
potential $V\left( g\right) $ for the scalar matter self-interaction, which
must allow for the spontaneous symmetry breaking mechanism.

Before specifying the model, let us recall that the main purpose of this
work is to investigate the existence of coherent twinlike models in the
context of classical Maxwell-Higgs field theories. In particular, we focus
our attention on the study of twinlike BPS states, i.e., 
finite energy field configurations which can be described as static
solutions of a set of first-order differential equations (the BPS equations)
having the very same field profile and energy.

In this sense, it is instructive to note that the limit $w\left( g\right) =1$
leads the model \eqref{1} back to the model studied in Ref.~\cite{b2}. Thus,
the limit $h\left( g\right) =1$ leads us back to the standard
Maxwell-Higgs electrodynamics. Thus, a very interesting choice to
the symmetry breaking potential $V\left( g\right) $ for the scalar matter
self-interaction is%
\begin{equation}
V_{s}\left( g\right) =\frac{1}{2}\left( g^{2}-1\right) ^{2}\text{ .}
\label{9}
\end{equation}%
Now, according to the conventions stated above, the 
Euler-Lagrange equations of motion \eqref{7} and \eqref{8} for the 
rotationally symmetric solutions can be rewritten in the form%
\begin{equation}
\frac{d^{2}a}{dr^{2}}-\frac{1}{r}\frac{da}{dr}=2g^{2}a\text{ ,}  \label{10}
\end{equation}%
\begin{equation}
\frac{d^{2}g}{dr^{2}}+\frac{1}{r}\frac{dg}{dr}-\frac{a^{2}g}{r^{2}}=g\left(
g^{2}-1\right) \text{ .}  \label{11}
\end{equation}%
The equations \eqref{10} and \eqref{11} are solvable by solutions of the
two first-order differential equations
\begin{equation}
\frac{dg}{dr}=\pm \frac{ga}{r}\text{ ,}  \label{12}
\end{equation}%
\begin{equation}
\frac{1}{r}\frac{da}{dr}=\pm \left( g^{2}-1\right) \text{ .}  \label{13}
\end{equation}

The static solutions $g\left( r\right) $ and $a\left( r\right) $ to the
above first-order equations are the well-known
Bogomol'nyi-Prasad-Sommerfield (BPS) states related to the standard
Maxwell-Higgs electrodynamics. These states solve the rotationally symmetric
Euler-Lagrange equations of motion \eqref{10} and \eqref{11} by minimizing
the total energy of the resulting solutions; see equations \eqref{16d} and %
\eqref{16e} below.

If we use non-trivial choices for the functions $h\left( g\right) $ and $%
w\left( g\right) $, the resulting Euler-Lagrange equations \eqref{7} and %
\eqref{8} will be much more sophisticated than the usual Eqs.~\eqref{10} and %
\eqref{11}. In this sense, it seems useful to calculate the general
expression for the non-standard energy density, from which we can get
interesting insights related to the non-usual BPS states. This expression is%
\begin{equation}
\varepsilon =\frac{h}{2}\left( \frac{1}{r}\frac{da}{dr}\right) ^{2}+w\left(
\left( \frac{dg}{dr}\right) ^{2}+\frac{a^{2}g^{2}}{r^{2}}\right) +V\text{ .}
\label{14}
\end{equation}%
Here we note that the presence
of the functions $h\left( g\right) $ and $w\left(
g\right) $ makes it very hard to obtain the BPS configurations related to
the non-standard model \eqref{1}. However, the existence of
coherent twinlike BPS configurations, i.e., finite energy static solutions
of the first-order eqs.\eqref{12} and \eqref{13}, is still possible, and it
is closely constrained by two important assumptions concerning the functions 
$h\left( g\right) $, $w\left( g\right) $ and $V\left( g\right) $. The first
one is the following differential relation%
\begin{equation}
w=h+\frac{\left( g^{2}-1\right) }{2g}\frac{dh}{dg}\text{ ,}  \label{15}
\end{equation}%
which leads to a link between the functions $h\left( g\right) $ and $w\left(
g\right) $. The second constraint is%
\begin{equation}
V\left( g\right) =h\left( g\right) V_{s}\left( g\right) \text{ ,}  \label{16}
\end{equation}%
which defines the potential $V\left( g\right) $ of the non-standard model %
\eqref{1} as a product between the function $h\left( g\right) $ and the
self-dual potential \eqref{9} of the canonical Maxwell-Higgs model. As we
demonstrate below, via the constraints \eqref{15} and \eqref{16}, the
non-standard energy functional \eqref{14} can be rewritten in a way such
that the corresponding total energy is minimized by the first order
equations \eqref{12} and \eqref{13}. Furthermore, such minimum energy can be
adjusted to mimic the very same value achieved by the standard BPS solutions.
This adjustment can be done by tuning the boundary conditions to be imposed
on $h\left( g\right) $, near the origin and asymptotically. We perform this
tuning in the next Section.

Now, to search for twinlike BPS solutions to the first-order equations %
\eqref{12} and \eqref{13}, we need to know the boundary conditions to be
imposed on the functions $g\left( r\right) $ and $a\left( r\right) $, near
the origin and asymptotically. Near the origin, such functions must have
no singularity. So, given the \textit{Ansatz} \eqref{5} and %
\eqref{6}, $g\left( r\right) $ and $a\left( r\right) $ have to behave
according to%
\begin{equation}
g\left( r\rightarrow 0\right) \rightarrow 0\text{ \ \ and \ \ }a\left(
r\rightarrow 0\right) \rightarrow n\text{ .}  \label{17}
\end{equation}%
Also, the twinlike BPS configurations must have finite total energy. In
this case, as a condition to make the energy finite, the non-standard energy
functional \eqref{14} must vanish for $r\rightarrow \infty $. So,
asymptotically, the profile functions $g\left( r\right) $ and $a\left(
r\right) $ must behave according to%
\begin{equation}
g\left( r\rightarrow \infty \right) \rightarrow 1\text{ \ \ and \ \ }a\left(
r\rightarrow \infty \right) \rightarrow 0\text{ .}  \label{18}
\end{equation}


\section{Twinlike BPS models}

\label{numerical}

Let us now focus our attention on the twinlike BPS configurations themselves.
The main purpose here is to use the theoretical
framework developed in the previous Section to assure the existence of such
twinlike configurations. We start using
\eqref{15} and \eqref{16} to rewrite the energy density \eqref{14} in a
way such that the resulting total energy is minimized by the differential eqs.~\eqref{12}
and \eqref{13}. The value achieved by such minimized energy
depends on the behaviour of $h\left( g\right) $, near the origin and
asymptotically. In this sense, by choosing appropriate boundary conditions,
we tune this value to be the very same one gets in the context of the
standard self-dual Maxwell-Higgs model \eqref{9}.

In general, since the functions $h\left( g\right) $ and $w\left( g\right) $
are related to each other by the differential constraint \eqref{15}, and the
non-standard potential $V\left( g\right) $ is given by Eq.~\eqref{16}, the
modified energy functional \eqref{14} can be rewritten in the form (see also
Eq.~\eqref{9})%
\begin{eqnarray}
\varepsilon &=&\frac{h}{2}\left( \frac{1}{r}\frac{da}{dr}\mp \left(
g^{2}-1\right) \right) ^{2}+w\left( \frac{dg}{dr}\mp \frac{ga}{r}\right) ^{2}
\notag \\
&&\pm \frac{1}{r}\frac{d}{dr}\left( \left( g^{2}-1\right) ah\right) \text{ ,}
\label{16a}
\end{eqnarray}%
from which we note that the corresponding total energy $E$ is minimized by
the first order differential equations \eqref{12} and \eqref{13}. In this
case, the minimum energy is given by%
\begin{equation}
E_{bps}=\int \varepsilon _{bps}d^{2}r\text{ ,}  \label{16b}
\end{equation}%
where%
\begin{equation}
\varepsilon _{bps}=\pm \frac{1}{r}\frac{d}{dr}\left( \left( g^{2}-1\right)
ah\right) \text{ ,}  \label{16c}
\end{equation}%
is the minimum energy density, i.e., the energy density related to the first
order equations \eqref{12} and \eqref{13}. Also, Eq.~\eqref{16b} defines the 
\textit{Bogomol'nyi bound}, i.e., the lower bound for the energy functional %
\eqref{14}.

We use Eq.~\eqref{16c} to see that the Bogomol'nyi bound \eqref{16b}
depends on the boundary conditions \eqref{17} and \eqref{18},
and also on the boundary conditions to be imposed on $h\left( g\right) $
itself. Here, the interesting point is that there are infinitely many
functional forms to $h\left( g\right) $ which obey the same general boundary
conditions; as a consequence, there are infinitely many non-standard energy
densities (all of them given by Eq.~\eqref{16c}) which achieve the very same
Bogomol'nyi bound (given by Eq.~\eqref{16b}). Each one of these forms of $%
h\left( g\right) $ defines a different non-standard self-dual model. In this
sense, since the first order equations \eqref{12} and \eqref{13}, and also
the boundary conditions \eqref{17} and \eqref{18}, do not depend on $h\left(
g\right) $, all these models exhibit the very same BPS\ solutions to
the profile functions $g\left( r\right) $ and $a\left( r\right) $. Since these
different models have the very same field configuration and
total energy, we keep naming them twinlike models.

To better clarify how the theoretical framework developed in this paper
works, let us first consider the standard Maxwell-Higgs model, which is
defined by $h\left( g\right) =1$. In such context, due to the finite-energy
boundary conditions \eqref{17} and \eqref{18}, Eq.~\eqref{16b} gives the
Bogomol'nyi bound (which is quantized according to the vorticity $n$ of the
solutions)%
\begin{equation}
E_{bps}=2\pi \left\vert n\right\vert \text{ ,}  \label{16d}
\end{equation}%
where%
\begin{equation}
\varepsilon _{bps}=\mp \frac{1}{r}\frac{da}{dr}\pm \frac{1}{r}\frac{d}{dr}%
\left( g^{2}a\right) \text{ ,}  \label{16e}
\end{equation}%
in the minimum energy functional. Here, we point out the existence of
infinitely many models (i.e., infinitely many functional forms to $h\left(
g\right) $) which achieve the very same Bogomol'nyi bound \eqref{16b}. In
fact, as the reader can easily check, any function $h\left( g\right) $
satisfying the boundary conditions ($C$ is any non-negative constant)%
\begin{equation}
h\left( r\rightarrow 0\right) \rightarrow 1\text{ \ \ and \ \ }h\left(
r\rightarrow \infty \right) \rightarrow C  \label{16f}
\end{equation}%
leads to $E_{bps}=2\pi \left\vert n\right\vert $. 

In order to illustrate the above result, let us now consider some specific models.
The first example, representing the first non-standard self-dual
model, is defined by%
\begin{equation}
h\left( g\right) =g^{2}+1\text{ .}  \label{16g}
\end{equation}%
In this case, $C=2$. The energy functional \eqref{16c} can be
rewritten as%
\begin{equation}
\varepsilon _{bps}=\mp \frac{1}{r}\frac{da}{dr}\pm \frac{1}{r}\frac{d}{dr}%
\left( g^{4}a\right) \text{ .}  \label{16h}
\end{equation}%
Then, via eqs.~\eqref{16b} and \eqref{16h}, we conclude that the modified
model defined by \eqref{16g} achieves the same Bogomol'nyi bound \eqref{16d}
achieved by the usual Maxwell-Higgs model.

The other non-standard self-dual model we present is defined by%
\begin{equation}
h\left( g\right) =\left( g^{2}+1\right) \left( g^{2}-1\right) ^{2}\text{ ,}
\label{16i}
\end{equation}%
with $C=0$ in this case. Here, the minimum energy density is%
\begin{equation}
\varepsilon _{bps}=\mp \frac{1}{r}\frac{da}{dr}\pm \frac{1}{r}\frac{d}{dr}%
\left( ag^{2}\left( g^{6}-2g^{4}+2\right) \right) \text{ .}  \label{16j}
\end{equation}%

It is important to note that, due to the boundary conditions \eqref{17}%
, \eqref{18} and \eqref{16f}, the second term in eqs.~\eqref{16e}, %
\eqref{16h} and \eqref{16j} does not contribute to the integration process in %
\eqref{16b}. In this sense, we conclude that
all the physical information concerning the Bogomol'nyi bound \eqref{16d}
achieved by the models previously presented is enclosed by the first term in
eqs.~\eqref{16e}, \eqref{16h} and \eqref{16j}. So, from now
on we refer to such first and second terms as the \textit{physical} and the 
\textit{irrelevant} ones, respectively. In fact, all models defined by a $%
h\left( g\right) $ of the form%
\begin{equation}
h\left( g\right) =1+G\left( g\right)  \label{gf}
\end{equation}%
exhibit a minimum energy functional given by%
\begin{equation}
\varepsilon _{bps}=\mp \frac{1}{r}\frac{da}{dr}\pm \frac{1}{r}\frac{d}{dr}%
\left( ag^{2}\left( 1+\frac{g^{2}-1}{g^{2}}G\right) \right) \text{ .}
\label{gf1}
\end{equation}%
Then, as a consequence, such models possess the very same physical term,
while exhibiting quite different irrelevant terms.

To end this Section, we point out that one can choose the boundary
conditions \eqref{16f} in a different way. Generically, one can define ($N$
is any positive constant)%
\begin{equation}
h\left( r\rightarrow 0\right) \rightarrow N\text{ \ \ and \ \ }h\left(
r\rightarrow \infty \right) \rightarrow C\text{ .}  \label{16l}
\end{equation}%
Here also, there are infinitely many functional forms to $h\left( g\right) $
which behave according to such boundary conditions. Each one of these forms
defines a different non-standard self-dual model. These models compose an
infinity set of twinlike theories, since they exhibit the very same BPS
configurations (given as solutions of the differential eqs.~\eqref{12} and %
\eqref{13}, according to the finite-energy boundary conditions \eqref{17}
and \eqref{18}) and also achieve the very same Bogomol'nyi bound (given by
Eq.~\eqref{16b}), which is%
\begin{equation}
E_{bps}=2\pi \left\vert n\right\vert N\text{ .}  \label{16m}
\end{equation}%
If we use a general $N$, none of the above models encloses the standard Maxwell-Higgs
model, which is defined with $N=1$. We also note that the extended versions of \eqref{gf} and %
\eqref{gf1} are quite obvious.


\section{Twinlike numerical solutions}
\label{numerical2}

Let us now perform the numerical study of the twinlike
self-dual models presented in this work. We numerically solve the first-order
differential equations \eqref{12} and \eqref{13}, according to the
finite energy boundary conditions \eqref{17} and \eqref{18}. The resulting
solutions are well-known, and can be found, for instance, in Ref.~\cite{n5}.
We then use such solutions to perform a detailed analysis of the
energy of the vortices, including their minimum energy densities %
\eqref{16e}, \eqref{16h} and \eqref{16j}. In this sense, we also depict,
separately, the numerical solutions to the physical and to the irrelevant
terms presented in the previous Section, from which we show (numerically) that
the irrelevant terms do not contribute to the Bogomol'nyi bounds computed above.
Some results are plotted below, for the case $n=1$.

In Fig.~1, we present the numerical solutions for the energy densities
related to the twinlike BPS states, and we note that the solutions for %
\eqref{16e} and \eqref{16j} engender the same basic features: they reach
their maximum values near the origin, and they decrease monotonically
for all $r$. Here, it is interesting to note that the core of the twinlike
solution \eqref{16j} is smaller than that of the usual solution \eqref{16e}.
On the other hand, we see that the solution for the minimum energy density %
\eqref{16h} is quite different from the previous ones. In this case, such
solution reaches its maximum value at some finite distance $R$ from the
origin, and it is not monotonically decrescent for all values of the
independent variable $r$. Finally, we point out that all the minimum energy
densities previously depicted vanish asymptotically.

\begin{figure}[tbph]
\centering\includegraphics[width=8.5cm]{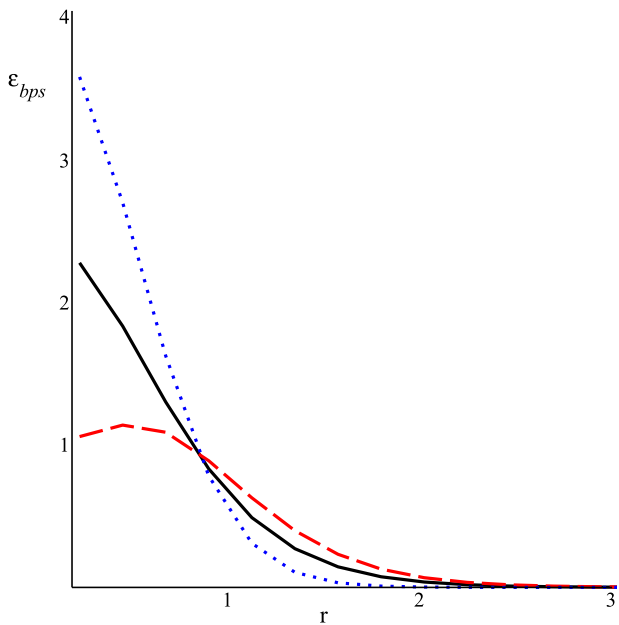}
\par
\vspace{-0.3cm}
\caption{Plots of solutions to the minimum energy densities \eqref{16e}
(black solid line), \eqref{16h} (red dashed line) and \eqref{16j} (blue
dotted line).}
\end{figure}

In fact, since the minimum non-standard energy functional \eqref{16c} is $r$%
-dependent only, the information regarding the Bogomol'nyi bound \eqref{16b}
is enclosed by the product $r\varepsilon _{bps}$. In particular, the total
energy $E_{bps}$ of the twinlike BPS states is given by $2\pi $ times the
area enclosed by $r\varepsilon _{bps}$. Regarding the
non-standard models previously presented, such area is always equal to $%
\left\vert n\right\vert $. Thus, all the above models achieve the
very same minimum total energy, which is given by the Bogomol'nyi bound %
\eqref{16d}.

\begin{figure}[tbph]
\centering\includegraphics[width=8.5cm]{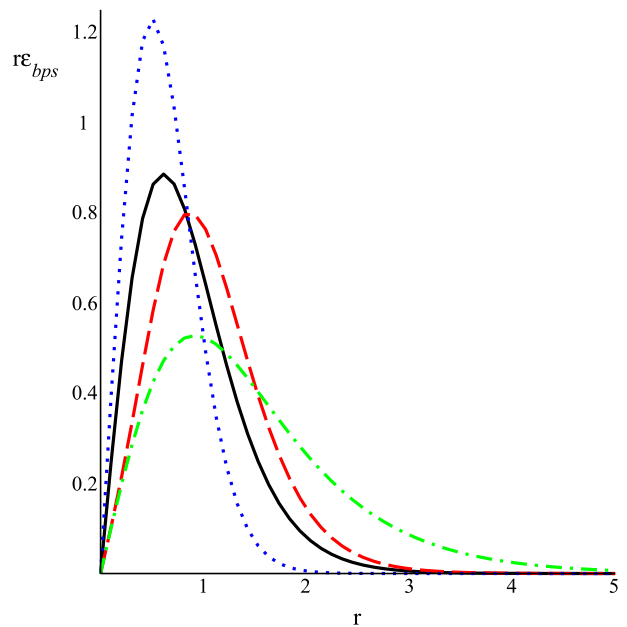}
\par
\vspace{-0.3cm}
\caption{Plots of solutions to the product $r\protect\varepsilon _{bps}$.
Conventions as in FIG. 1. See also the solution to $-a^{\prime }$ (green
dot-dashed line).}
\end{figure}

In Fig.~2, we depict the numerical solutions for the product $r\varepsilon
_{bps}$ related to the previous models. In general, all the solutions
engender the same features: they reach their maximum values at some finite
distance from the origin, while vanishing for $r\rightarrow 0$ and for $%
r\rightarrow \infty $. It is interesting to note how these
solutions behave in order to enclose the same area ever: there is an inverse
relation between their maximum values and their characteristic lengths, and,
as a consequence, the solutions with greater amplitudes exhibit smaller
cores, and vice-versa.

In fact, the product $r\varepsilon _{bps}$ means a summation over two
different terms: the first one is given by $r$ times the physical term, and
the second one is given by $r$ times the irrelevant term; see eqs.\eqref{16e}%
, \eqref{16h} and \eqref{16j}. So, in order to get to some useful insights
concerning the total energies of the twinlike BPS states, let us evaluate
the numerical solutions for such terms, separately.

\begin{figure}[tbph]
\centering\includegraphics[width=8.5cm]{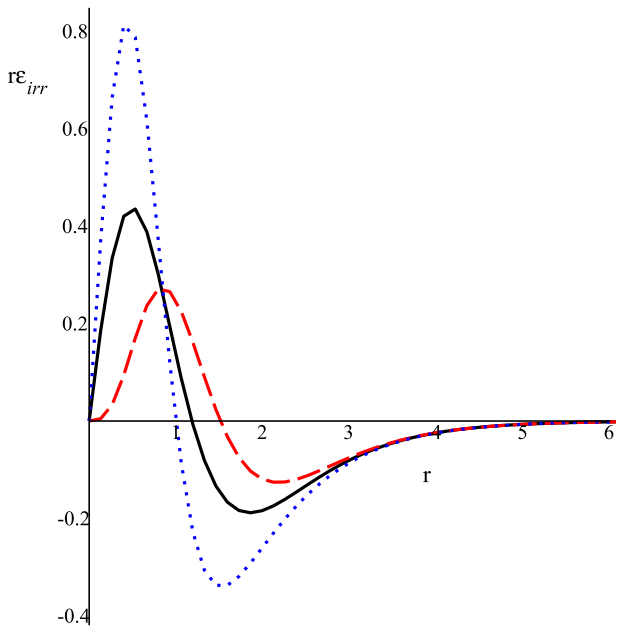}
\par
\vspace{-0.3cm}
\caption{Plots of solutions to the product between $r$ and the irrelevant
terms. Conventions as in FIG. 1.}
\end{figure}

In the same Fig.~2, we plot the solution to $-a^{\prime }$ (prime means a
derivative with respect to $r$). This solution is important since it denotes
the product between $r$ and the physical term, and we note that it exhibits
the same basic features the solutions to $r\varepsilon _{bps}$ do, including
the inverse relation between the amplitude and the characteristic length.
Here, it is interesting to note that this solution also encloses an area
equal to $\left\vert n\right\vert $. So, as stated theoretically before, our
numerical analysis reinforces that all the physical information regarding
the total energies of the non-standard models previously presented is
enclosed by the physical term itself.

In Fig.~ 3, we depict the solutions to the second term, that is, the
product between $r$ and the irrelevant term in the energy density. 
In this case, it is interesting
to note how these solutions behave in order to contribute nothing to the
Bogomol'nyi bound. Here, we reinforce that, even in the presence of the
negative sectors presented by such solutions (and, as a consequence, by the
irrelevant terms themselves, since the radial coordinate $r$ is always
positive), the positiveness of the minimum energy functionals \eqref{16e}, %
\eqref{16h} and \eqref{16j} is completely assured; see, for instance, the
solutions depicted in Fig. 1.

To end this Section, let us investigate a very important issue regarding the
study of topological configurations, which is the existence of conserved
topological charges. In fact, the existence of such charges assures the
topological stability of the corresponding configurations. In the present
context, that is, for rotationally symmetric solutions of the form \eqref{5}
and \eqref{6}, the corresponding topological charge $Q_{T}$ can be
identified with the flux of the magnetic field generated by such solutions
themselves. So, one has%
\begin{equation}
Q_{T}=\int Bd^{2}r\equiv \Phi _{B}\text{ ,}  \label{36}
\end{equation}%
where $\Phi _{B}$ stands for the magnetic flux, and%
\begin{equation}
B\left( r\right) =-\frac{1}{r}\frac{da}{dr}
\end{equation}%
is the magnetic field; see \eqref{6}. According to our conventions, the
magnetic flux (and, as a consequence, the topological charge) can be written
as%
\begin{equation}
\Phi _{B}=2\pi n\text{ ,}  \label{37}
\end{equation}%
which shows that $Q_{T}$ is conserved, and is quantized according to the
vorticity $n$. Since this result does not depend on $h\left( g\right) $,
we think that the topological stability of the non-standard solutions is
achieved in the same way as that of the standard case.


\section{Ending comments}

\label{end}

In this work we investigated the existence of consistent twinlike self-dual
classical field theories in the context of the planar Maxwell-Higgs model.
We have used a modified Maxwell-Higgs model endowed by non-standard dynamics
to both the gauge and the scalar fields. The unusual dynamics
were introduced in terms of two non-trivial functions, $h\left( \left\vert
\phi \right\vert \right) $ and $w\left( \left\vert \phi \right\vert \right) $, 
which are functions of the amplitude of the scalar field; they must be
positive, in order to assure the positiveness of the energy functional of
the modified model. We note that $h\left( \left\vert \phi \right\vert
\right) $ is a kind of \textit{generalized dielectric function} which couples to the
canonical Maxwell action, leading to a non-standard
dynamics to the gauge field. Also, $w\left( \left\vert \phi \right\vert
\right) $ couples to the squared covariant derivative of the scalar field.

We have demonstrated the existence of twinlike
self-dual Abelian-Higgs models under the presence of two specific
constraints: the first one is a differential relation between the
non-trivial functions $h\left( \left\vert \phi \right\vert \right) $ and $%
w\left( \left\vert \phi \right\vert \right) $, and the second one is a
relation between $h\left( \left\vert \phi \right\vert \right) $ and the
spontaneous symmetry breaking potential $V\left( \left\vert \phi \right\vert
\right) $ of the non-standard model. We have used these constraints to performed a detailed
numerical investigation, studying the energy spectra of the generalized models,
including their minimum energy functionals. In general,
we have noted that the numerical profiles for $\varepsilon _{bps}$ itself
can be quite different from one another, as they can engender different
basic features. Even in this case, we have noticed that the solutions to $%
r\varepsilon _{bps}$ engender the same features, including an interesting
inverse relation between their maximum values and their characteristic
lengths. In particular, such relation is very important, since it assures
that all the solutions to $r\varepsilon _{bps}$ enclose the same area, so
all the non-standard BPS states have the very same
total energy.

We hope that the above results may stimulate subsequent analysis in the field,
mainly regarding the features that the twinlike Maxwell-Higgs models may
engender. In particular, a rather natural issue concerns the extension of
the present idea to the case of non-rotationally symmetric solutions. Another
issue concerns the extension of the twinlike models to the context of Yang-Mills-Higgs theories.
These and other related issues are under investigation, and we hope to report on them in the
near future.

The authors would like to thank CAPES, CNPq (Brazil) and FCT Project
CERN/FP/116358/2010 (Portugal) for partial financial support. E. da Hora
thanks the Department of Mathematical Sciences of Durham University (U.K.),
for all their hospitality while doing this work.


\begin{thebibliography}{99}
\bibitem{n5} A. Vilenkin and E. P. S. Shellard, \textit{Cosmic Strings and
Other Topological Defects} (Cambridge University Press, Cambridge, England,
1994); N. Manton and P. Sutcliffe, \textit{Topological Solitons} (Cambridge
University Press, Cambridge, England, 2004).

\bibitem{n0} D. Finkelstein, J. Math. Phys. \textbf{7}, 1218 (1966).

\bibitem{n1} A. A. Abrikosov, Sov. Phys. JETP \textbf{5}, 1174 (1957); H. B.
Nielsen and P. Olesen, Nucl. Phys. B \textbf{61}, 45 (1973).

\bibitem{n3} G. 't Hooft, Nucl. Phys. B \textbf{79}, 276 (1974); A. M.
Polyakov, JETP Lett. \textbf{20}, 194 (1974).

\bibitem{Bogo} E. Bogomol'nyi, Sov. J. Nucl. Phys. \textbf{24}, 449 (1976);
M. Prasad and C. Sommerfield, Phys. Rev. Lett. \textbf{35}, 760 (1975).

\bibitem{todos} E. Babichev, Phys. Rev. D 74, 085004 (2006); Phys. Rev. D 
\textbf{77}, 065021 (2008); D. Bazeia, L. Losano, R. Menezes and J. C. R. E.
Oliveira, Eur. Phys. J. C \textbf{51}, 953 (2007); X. Jin, X. Li and D. Liu,
Classical Quantum Gravity \textbf{24}, 2773 (2007); C. Adam, N. Grandi, J.
Sanchez-Guillen and A. Wereszczynski, J. Phys. A \textbf{41}, 212004 (2008);
Erratum-ibid. A \textbf{42}, 159801 (2009); C. Adam, J. Sanchez-Guillen and
A. Wereszczynski, J. Phys. A \textbf{40}, 13625 (2007); Erratum-ibid. A. 
\textbf{42}, 089801 (2009); C. Adam, N. Grandi, P. Klimas, J.
Sanchez-Guillen and A. Wereszczynski, J. Phys. A \textbf{41}, 375401 (2008);
C. Adam, P. Klimas, J. Sanchez-Guillen and A. Wereszczynski, J. Phys. A 
\textbf{42}, 135401 (2009); D. Bazeia, E. da Hora, C. dos Santos and R.
Menezes, Phys. Rev. D \textbf{81}, 125014 (2010); D. Bazeia, E. da Hora, R.
Menezes, H. P. de Oliveira and C. dos Santos, Phys. Rev. D \textbf{81},
125016 (2010); C. dos Santos and E. da Hora, Eur. Phys. J. C \textbf{70},
1145 (2010); Eur. Phys. J. C \textbf{71}, 1519 (2011); C. dos Santos, Phys.
Rev. D \textbf{82}, 125009 (2010).

\bibitem{ALTW} M. Andrews, M. Lewandowski, M. Trodden and D. Wesley, Phys.
Rev. D \textbf{82}, 105006 (2010).

\bibitem{ufpb} D. Bazeia, J. D. Dantas, A. R. Gomes, L. Losano and R.
Menezes, Phys. Rev. D \textbf{84}, 045010 (2011).

\bibitem{adam} C. Adam and J. M. Queiruga, Phys. Rev. D \textbf{84}, 105028
(2011).

\bibitem{bm} D. Bazeia and R. Menezes, Phys. Rev. D \textbf{84}, 125018
(2011).

\bibitem{quarks} R. Friedberg and T. D. Lee, Phys. Rev. D \textbf{15}, 1694
(1977); Phys. Rev. D \textbf{16}, 1096 (1977); Phys. Rev. D \textbf{18},
2623 (1978).

\bibitem{b2} D. Bazeia, Phys. Rev. D \textbf{46}, 1879 (1992).
\end{thebibliography}
\end{document}